\documentclass[epj]{webofc}
\usepackage[utf8]{inputenc}
\usepackage[varg]{txfonts}   
\usepackage{booktabs}
\usepackage{xcolor}
\definecolor{darkred}{rgb}{0.4,0.0,0.0}
\definecolor{darkgreen}{rgb}{0.0,0.4,0.0}
\definecolor{darkblue}{rgb}{0.0,0.0,0.4}
\usepackage[bookmarks,linktocpage,colorlinks,
    linkcolor = darkred,
    urlcolor  = darkblue,
    citecolor = darkgreen]{hyperref}
\usepackage{subfigure}
\wocname{EPJ Web of Conferences}
\woctitle{Lattice2017}
\newcommand{\bmat}{\left ( \begin{array}{cc}}
\newcommand{\emat}{ \end{array}\right )}

\renewcommand\epsilon\varepsilon
\renewcommand\phi\varphi

\newcommand\be{\begin{eqnarray}}
\newcommand\ee{\end{eqnarray}}

\newcommand{\Tr}{{\rm Tr}}

\DeclareMathOperator{\tr}{tr}

\begin{document}


%
\selectlanguage{english}
\title{%
Progress on Complex Langevin simulations of a finite density matrix model for QCD
}
\author{%
\firstname{Jacques} \lastname{Bloch}\inst{1} \and
\firstname{Jonas} \lastname{Glesaaen}\inst{2}\and
\firstname{Jacobus}  \lastname{Verbaarschot}\inst{3}\and\firstname{Savvas} \lastname{Zafeiropoulos}\inst{4,5,6}\fnsep\thanks{Speaker, \email{savvas@jlab.org} }
}
\institute{%
Institute for Theoretical Physics, University of Regensburg, 93040 Regensburg, Germany
\and
Department of Physics, College of Science, Swansea University,
Swansea, SA2 8PP, United Kingdom
\and
Department of Physics and Astronomy, Stony Brook University, Stony Brook, NY, 11794 
\and 
Thomas Jefferson National Accelerator Facility, Newport News, VA 23606, USA\and
Department of Physics, College of William and Mary, Williamsburg, VA 23187-8795, USA\and
Institute for Theoretical Physics, Universit\"at Heidelberg, Philosophenweg 12, D-69120 Germany
}
\abstract{%
 We study the Stephanov model, which is an RMT model for QCD at finite density, using the Complex Langevin algorithm. Naive implementation of the algorithm shows convergence towards the phase quenched or quenched theory rather than to intended theory with dynamical quarks. A detailed analysis of this issue and a potential resolution of the failure of this algorithm are discussed. 
We study the effect of gauge cooling on the Dirac eigenvalue distribution and time evolution of the norm for various cooling norms, which were specifically designed to remove the pathologies of the complex Langevin evolution. The cooling is further supplemented with a shifted representation for the random matrices. Unfortunately, none of these modifications generate a substantial improvement on the complex Langevin evolution and the final results still do not agree with the analytical predictions.
}
\maketitle
\section{Introduction}\label{intro}

Many interesting physical systems have a complex action which impedes Monte
Carlo simulations due to the notorious sign problem. In this category belongs
QCD at finite baryon density, Yang-Mills theories in the presence of a
$\theta$-term, systems of strongly correlated electrons and many other 
interesting physical systems. The Complex Langevin (CL) algorithm has been
advocated as one of the most promising routes for a solution of
the sign problem \cite{Parisi:1984cs,Aarts:2008wh}.
Despite some successes in toy models, the algorithm seems to
have  pathological issues such as converging to the wrong theory.
Detailed studies have also revealed that the criteria which
were put forward in order to guarantee a correct result are not fulfilled in
practice in many cases of interest such as cold and dense QCD close to the
chiral limit. An update of the current status can be found in~\cite{ES}.  In this
talk we discuss a Random Matrix Theory (RMT) model of QCD at nonzero baryon
density, which has an exponentially hard sign problem, but serves as an excellent
test bed for new algorithms since it can be solved analytically. This model
is based on a model for QCD at nonzero nonzero temperature
or imaginary chemical potential \cite{Jackson:1995nf} and was extended
to QCD at nonzero chemical potential  by Stephanov in~\cite{Misha}. It has a rich phenomenological
structure, in particular,  it shows a first order transition to a phase
of nonzero baryon density.
RMT has provided a great deal of analytical results for
non-perturbative aspects of QCD, including finite density, lattice cutoff
effects and topology~\cite{DSV,ADSV,KVZprl,KVZprd,KVZ2c,CSZ, pdrmt,like}.
First results of our studies were presented in~\cite{conf16}.

\section{Random Matrix Model}
We study an RMT model for QCD at finite density,  proposed by Stephanov \cite{Misha}. This model is defined by its partition function
 \begin{equation}
 \mathcal{Z}^{N_f}_N=e^{N\mu^2}\int dW dW^{\dagger}{\det}^{N_f}(D+m) \, e^{-N\,\Tr WW^\dagger}.
 \label{Zst}
 \end{equation}
 Originally this model at  imaginary chemical potential was introduced
 as a model for QCD at nonzero temperature \cite{Jackson:1995nf} in which case it has
 a second order chiral symmetry restoration transition.
  The block structure of the Dirac operator $D$ is
 \begin{equation}
  D=\left(
 \begin{array}{cc}
 m        & iW+\mu\\
 iW^\dagger+\mu & m
 \end{array}
 \right),
 \label{Dsteph}
\end{equation}
where a term proportional to the baryon chemical potential $\mu\gamma_0$ has
been added to the chRMT Dirac operator first proposed in \cite{ShVe93}. The
matrix $W$ is a random complex matrix of size $N\times(N+\nu)$, consisting of
Gaussian distributed random numbers, $N$ is the size of the block matrix $W$, and
$\nu$ is the analogue of the topological charge. Here we focus on the  $\nu=0$
case, because the topological charge has a negligible effect  on the observables that
we study.  In direct analogy to QCD, RMT simulations are similarly hampered by the
sign problem, which becomes exponentially hard when the quark mass is inside the
support of the Dirac spectrum.

We mainly focus on two  observables, the mass dependent chiral condensate   $\Sigma(m)=\langle\bar{\eta}\eta\rangle$
and the baryon number density $n_B=\langle \eta^{\dagger}\eta\rangle$
which are defined as 
\begin{equation}
\langle\bar{\eta}\eta\rangle=\frac 1{2N}\frac{\partial \log{ \mathcal{Z}^{N_f}_N}}{\partial m},
\label{PBP} \qquad \mbox{ and  } \qquad \langle\eta^{\dagger}\eta\rangle=\frac 1{2N}\frac{\partial \log{ \mathcal{Z}^{N_f}_N}}{\partial \mu}.
\end{equation}
In order to understand better the success and failure of the complex Langevin
algorithm we will, in parallel to the Stephanov model, also study a similar matrix model for QCD at finite density which turns out not to have a transition to a phase with nonzero baryon density. This is the Osborn
model~\cite{James}, which is a model for QCD at low baryon density. It is a
two-matrix model which is structurally quite similar to the Stephanov model and
its partition function reads
\begin{equation}
 \mathcal{Z}^{N_f}_N=e^{N\mu^2}\int dWdW' dW^{\dagger}dW'^{\dagger}{\det}^{N_f}(D+m)e^{-N\,\Tr (WW^\dagger+W'W'^\dagger)}.
 \label{Zosb}
\end{equation}
The Dirac operator $D$ has the form
\begin{equation}
  D=\left(
 \begin{array}{cc}
 m        & iW+\mu W'\\
 iW^\dagger+\mu W'^{\dagger} & m
 \end{array}
 \right).
 \label{Dsteph}
\end{equation}
However this model has a quite trivial dependence on the baryon chemical potential since its  partition function at $\mu \ne 0$ is related to the one at $\mu=0$ by a multiplicative factor, as follows \cite{James,Bloch:2012ye}
\begin{equation}\label{mufactorization}
\mathcal{Z}^{N_f}_N(m,\mu) = (1-\mu^2)^{N_f N}\mathcal{Z}^{N_f}_N\left(\frac{m}{\sqrt{1-\mu^2}},0\right).
\end{equation}
Due to the above factorization property of the partition function the Osborn model does not possess a phase transition to a phase with non-zero baryon density and thus can only describe properties of QCD for low values of the baryon chemical potential and is expected to have a simpler sign problem compared to the model of Stephanov. In this work, we will check in detail some of the proposed techniques which could in principle fix the problematic issues of the CL algorithm.

The partition function of the Stephanov model can be brought to a
form that allows for an easy numerical evaluation at finite $N$, or can be
treated completely analytically by a saddle point approximation when the matrix
size $N\to\infty$ \cite{HJV}.  For the  case of one dynamical flavor, $N_f=1$, the partition
function, in units where the chiral condensate $\Sigma=1$, can be reduced to a
$\sigma$-model via bosonization
\begin{eqnarray}
  \mathcal{Z}^{N_f=1}_N(m,\mu) = e^{N\mu^2}\int d\sigma d\sigma^* e^{-N\sigma^2} %
    (\sigma \sigma^* + m(\sigma+\sigma^*) + m^2 -\mu^2)^N \ ,
  \label{zoneflavor}
\end{eqnarray}
where $\sigma$ is the bosonized version of $\bar{\psi}_L \psi_R$. If one
changes variables to polar coordinates, the angular integral can be computed
analytically and it evaluates to a modified Bessel function. In the end, the
partition function becomes a one-fold integral
\begin{eqnarray}
  \mathcal{Z}^{N_f=1}_N(m,\mu) = \pi e^{-Nm^2+N\mu^2}\int_0^\infty du %
    \, (u-\mu^2)^N I_0(2mN\sqrt u) \, e^{-Nu} \ .
 \label{zi0}
\end{eqnarray}
We perform numerical simulations of the Stephanov model employing the Complex
Langevin algorithm and we test the algorithm  by comparing the numerical data
from our simulations to analytical results for the chiral
condensate and the baryon density that we can derive by differentiating the
partition function (\ref{zi0}) with respect to the corresponding sources, $m$ and $\mu$. We
also perform simulations of the Osborn model in order to better sketch how
the CL algorithm can handle models (or theories in general) that have a strong or a
mild sign problem.

\section{Implementation of the algorithm}

\begin{figure}[ht] 
  \begin{minipage}[b]{0.5\linewidth}
    \centering
    \includegraphics[width=.95\linewidth]{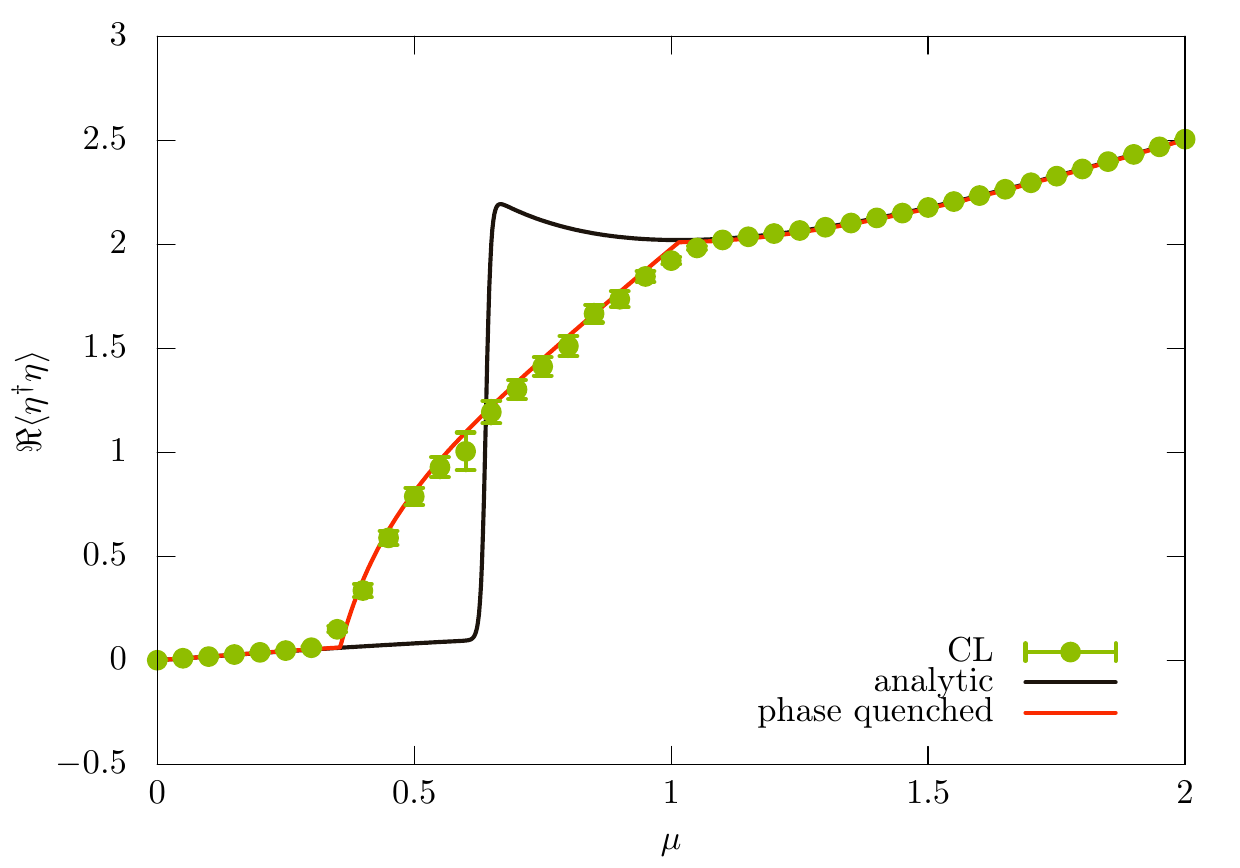} 
  \end{minipage}
  \begin{minipage}[b]{0.5\linewidth}
    \centering
    \includegraphics[width=.95\linewidth]{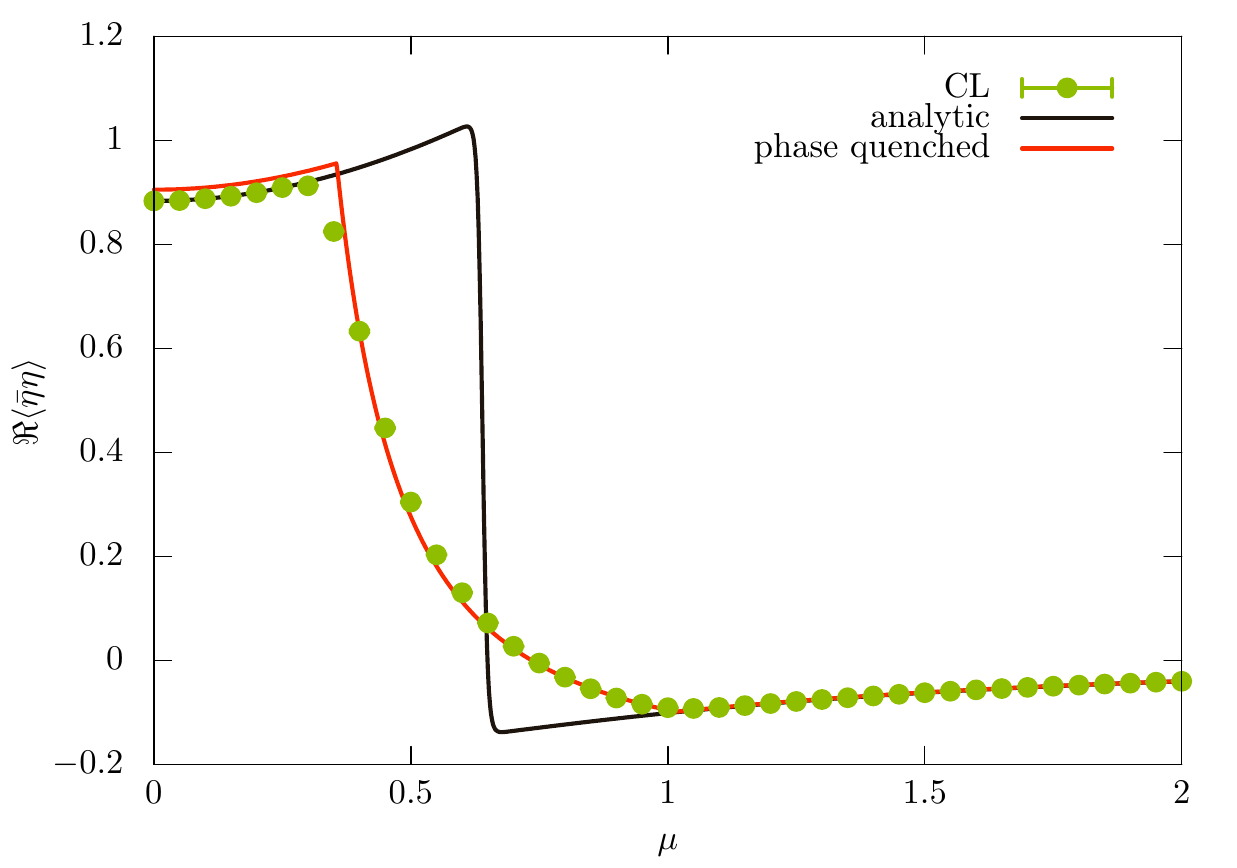} 
  \end{minipage} 
  \caption{ $\mu$-scan for $m=0.2$ and $N=48$. We show the baryon number
    density on the left and the chiral condensate on the right.}
  \label{muscan}
\end{figure}

When employing the CL algorithm, the real degrees of freedom of the original
model will take on complex values; this is called \emph{complexification}. To
clearly distinguish between pre- and post-complexification variables, we
introduce the following notation,
\begin{equation}
  W = X, \hspace{.5cm} \text{and} \hspace{.5cm} W^{\dagger} = Y,
\end{equation}
where $W$ describes the RMT system before complexification, while $X$ and $Y$ are general complex matrices describing the complexified system. 
At zero Langevin time, $\tau = 0$, we have $Y^{\dagger} = X$, but this will
no longer be the case for $\tau > 0$. 
%
In the Cartesian representation the complex matrices $X$ and $Y$ are given by
\begin{equation}
  X = A + i B, \hspace{.5cm} \text{and} \hspace{.5cm} Y = A^T - i B^T,
\end{equation}
where $\{A, B\} \in \mathbb{R}$ for $\tau = 0$, while $\{A, B\} \in \mathbb{C}$
for $\tau > 0$.  As we already showed in~\cite{conf16} a naive implementation of
the algorithm leads to convergence to the phase quenched theory, as is clearly
demonstrated in figure \ref{muscan}.

\section{Gauge cooling}

The RMT action is invariant under a $U(N)\times U(N+\nu)$ gauge transformation,
whereas after the variables have been complexified, this invariance is enhanced
to a $Gl(N)\times Gl(N+\nu)$ symmetry. One can utilize this additional gauge
symmetry in order to modify the Langevin evolution in the hope to retrieve the correct results for this model. 
This method of treating the pathologies of the
algorithm has been coined as  \emph{gauge cooling}, and has been 
employed with success on certain models \cite{sexty1, Nagata1, Nagata:2016mmh}. Closely
related to our study is its successful application to the Osborn
model~\cite{Nagata1}.

\subsection{Cooling norms}

The matrices $h\in Gl(N)$, so that $\{X, Y\} \to h \{X, Y\} h^{-1}$, are
constructed in a way that reduces a \emph{cooling norm}. Various norms are built
in ways that parametrize pathological features of the Complex Langevin evolution.
The natural norm is the Hermiticity norm which quantifies the difference
between $X^{\dagger}$ and $Y$,
\begin{equation}
  \mathcal{N}_H = \frac{1}{N} \tr \Big[ \big(X - Y^{\dagger}\big)^{\dagger} %
    \big(X - Y^{\dagger}\big) \Big].
\end{equation}
One can also introduce an eigenvalue norm \cite{Nagata1}
\begin{equation}
  \mathcal{N}_{\mathrm{ev}} = \sum_{i = 1}^{n_{\mathrm{ev}}} e^{-\xi \gamma_i},
\end{equation}
where $\gamma_i$ are the $n_{\mathrm{ev}}$ lowest eigenvalues of the positive definite matrix
$D^{\dagger} D$, and $\xi$ is a real positive parameter. Finally we can define an anti-Hermiticity norm which reads
\begin{equation}
  \mathcal{N}_{AH} %
    = \frac{1}{N} \tr \Big[\Big( \big(\phi + \psi^{\dagger}\big)^{\dagger} %
    \big(\phi + \psi^{\dagger}\big) \Big)^p\Big],
\end{equation}
The matrices $\psi$ and
$\phi$ are the off-diagonal elements of $D$: $\psi = i Y + \mu$,
$\phi = i X + \mu$.
The first two  norms as well as the anti-Hermiticity norm for $p=1$
were first proposed in \cite{Nagata1}. For the Stephanov model, we are
using the anti-Hermiticity norm with $p=2$ because for $p=1$ the
derivatives with respect to the similarity transformation are
independent of the chemical potential. For the Osborn model, where this
is not a problem, we use $p=1$.
Usually different norms attempt to fix different problems of the Langevin evolution so one can also combine them
to make aggregate norms. For example one could combine the Hermiticity norm,
which gives the magnitude of drift into the imaginary plane,
with either the anti-Hermiticity or the eigenvalue norm, which both deal with the singularities of the drift, %
\begin{equation}
  \mathcal{N}_{\mathrm{agg}} = (1-s) N_{AH/\mathrm{ev}} + s N_H,
  \hspace{.25cm}\text{where}\; s \in [0, 1].
\end{equation}
We refer the reader to ~\cite{Nagata1,CLarticle} for the details on how to construct the matrix $h$.

\subsection{Gauge cooling results}

Next, we present our results when applying the method of gauge cooling to the
two aforementioned matrix models. In these runs we have used a block matrix of size
$N = 24$, Langevin stepsize $\mathrm{d}t = 10^{-4}$, and runtime
$t_{\mathrm{end}} = 1$. Between each Langevin update we perform ten cooling
transformations.
 In what follows the Stephanov model is simulated at parameters
$\{m = 0.2, \mu = 0.5\}$, while we chose $\{m = 0.1, \mu = 0.25\}$ for the
Osborn model. Parameters are chosen in a way that the two models have a
seemingly equally severe sign problem to overcome.

\begin{figure}
  \begin{center}
    \includegraphics[width=0.45\textwidth]{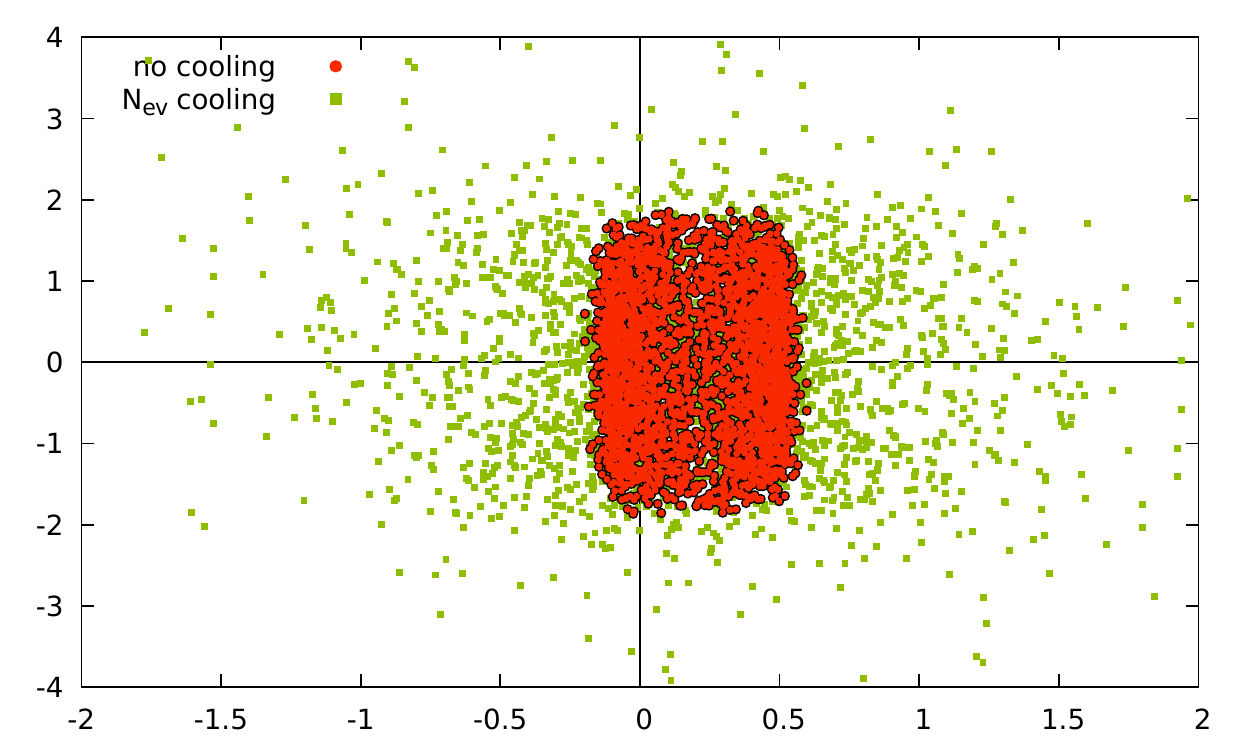}
    \includegraphics[width=0.45\textwidth]{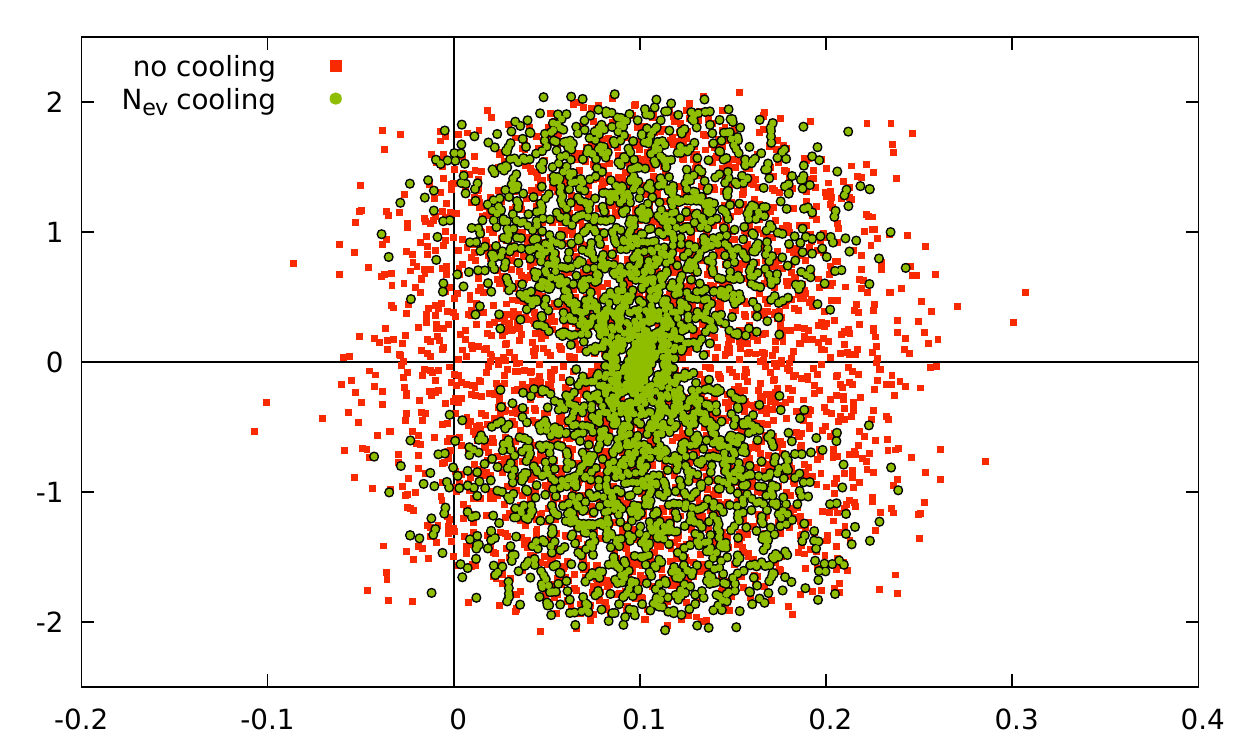}
  \end{center}
  \caption{Scatter plots of the eigenvalues of the fermion matrix for a standard
    CL run (red points) together with the ones from a gauge cooled run (green poits). We chose the
    parameters $\{\xi = 100, n_\mathrm{ev} = 2\}$ for
    $\mathcal{N}_{\mathrm{ev}}$. The plots show the eigenvalues from the last 60
    trajectories, separated by 100 updates. The left hand plot shows the
    Stephanov model, while the Osborn model is shown to the right.}
  \label{fig:eval_cooling_eigenvalues}
\end{figure}

\begin{figure}
  \begin{center}
    \includegraphics[width=0.45\textwidth]{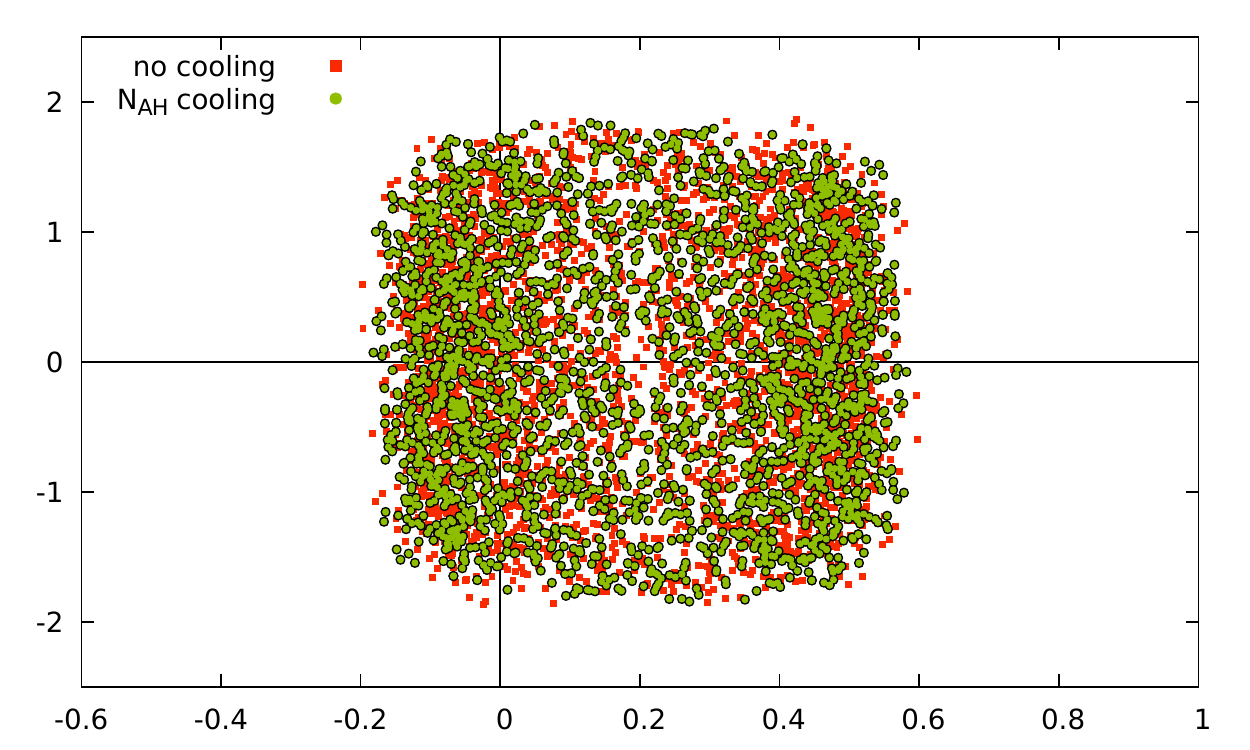}
    \includegraphics[width=0.45\textwidth]{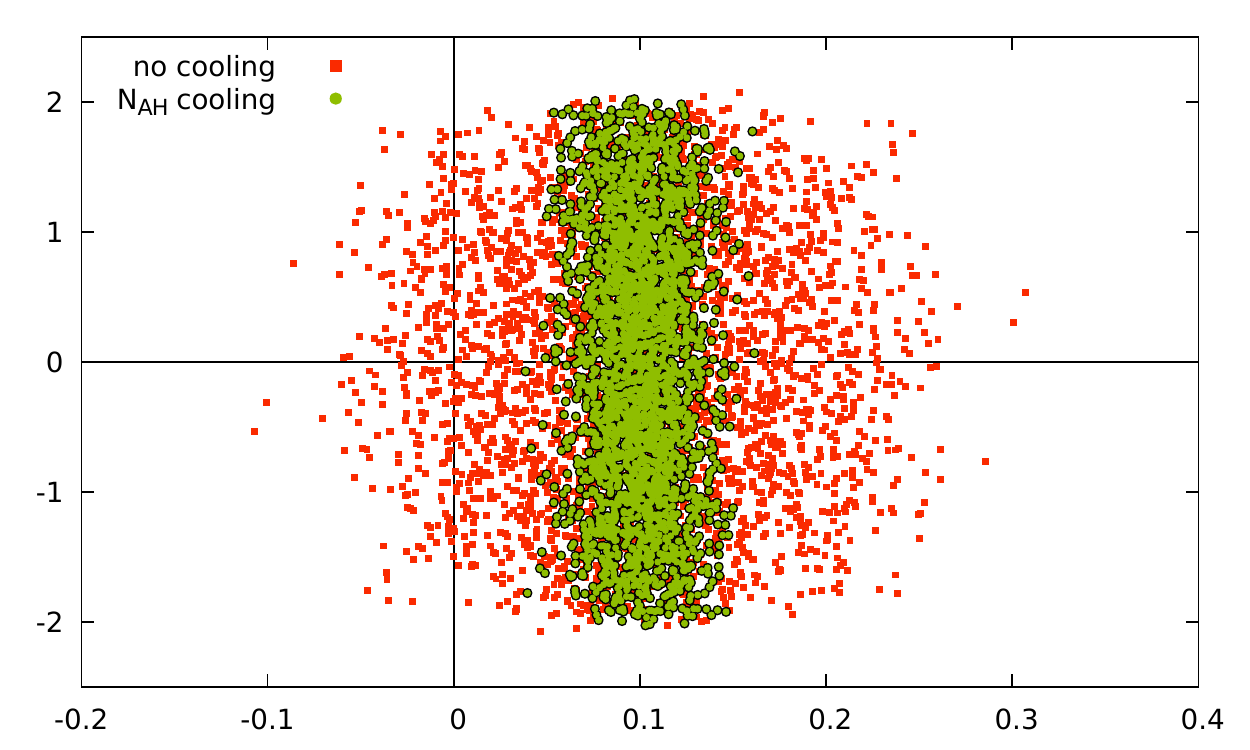}
  \end{center}
  \caption{Scatter plots of the Dirac eigenvalues for a standard
    CL run (red points) together with the ones from a run cooled with the $\mathcal{N}_{AH}$
    cooling norm (green points). The plots show the eigenvalues from the last 60 trajectories,
    separated by 100 updates. The LHS plot shows the Stephanov model,
    while the RHS shows the results of the Osborn model.}
  \label{fig:ah_cooling_eigenvalues}
\end{figure}

We start by looking at scatter plots of the eigenvalues of the Dirac matrix,
shown in  figure \ref{fig:eval_cooling_eigenvalues}. In direct analogy with the
results obtained by~\cite{Nagata:2016mmh} the RHS plot shows the effects of
cooling on the Osborn model. For this model, the effect of cooling
is that the eigenvalue
distribution deforms and avoids the origin. On the other hand, this is not the
case for the Stephanov model shown on the LHS.
Figure~\ref{fig:ah_cooling_eigenvalues} shows the same models cooled via
$\mathcal{N}_{AH}$. Here again, we observe that cooling cures the pathologies of
the Osborn model but not of the Stephanov model.

\begin{figure}
  \begin{center}
    \includegraphics[width=0.45\textwidth]{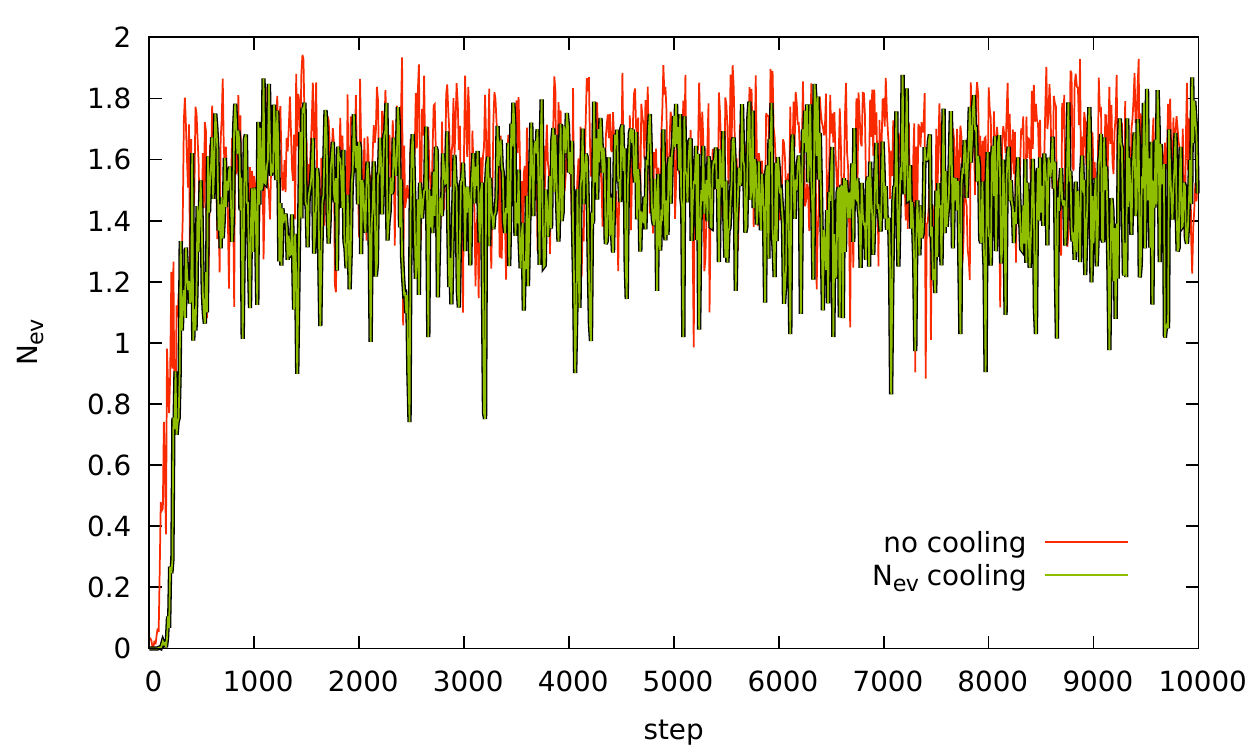}
    \includegraphics[width=0.45\textwidth]{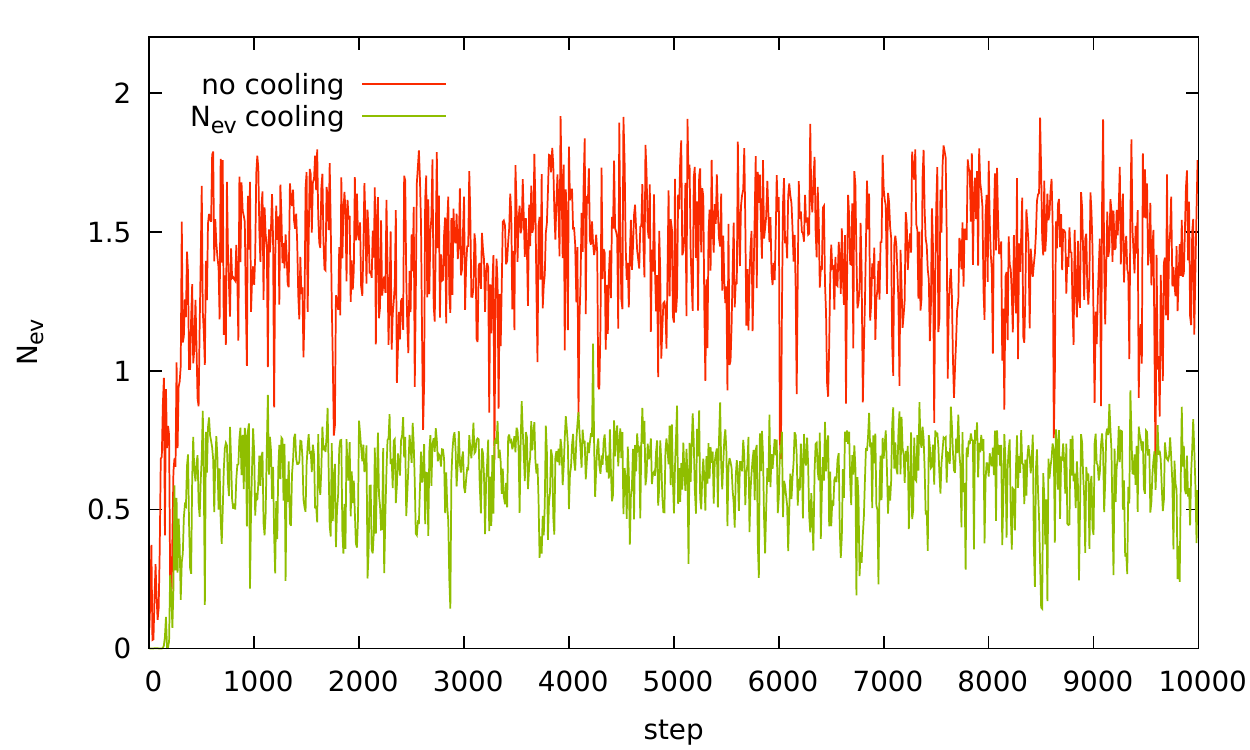}
  \end{center}
  \caption{Value of the eigenvalue norm, $\mathcal{N}_{\mathrm{ev}}$,
with (green curves) and without cooling (red curves)
    as a function of Langevin time.
    Stephanov model (LHS), Osborn model (RHS). }
  \label{fig:eval_cooling_norm}
\end{figure}

\begin{figure}
  \begin{center}
    \includegraphics[width=0.45\textwidth]{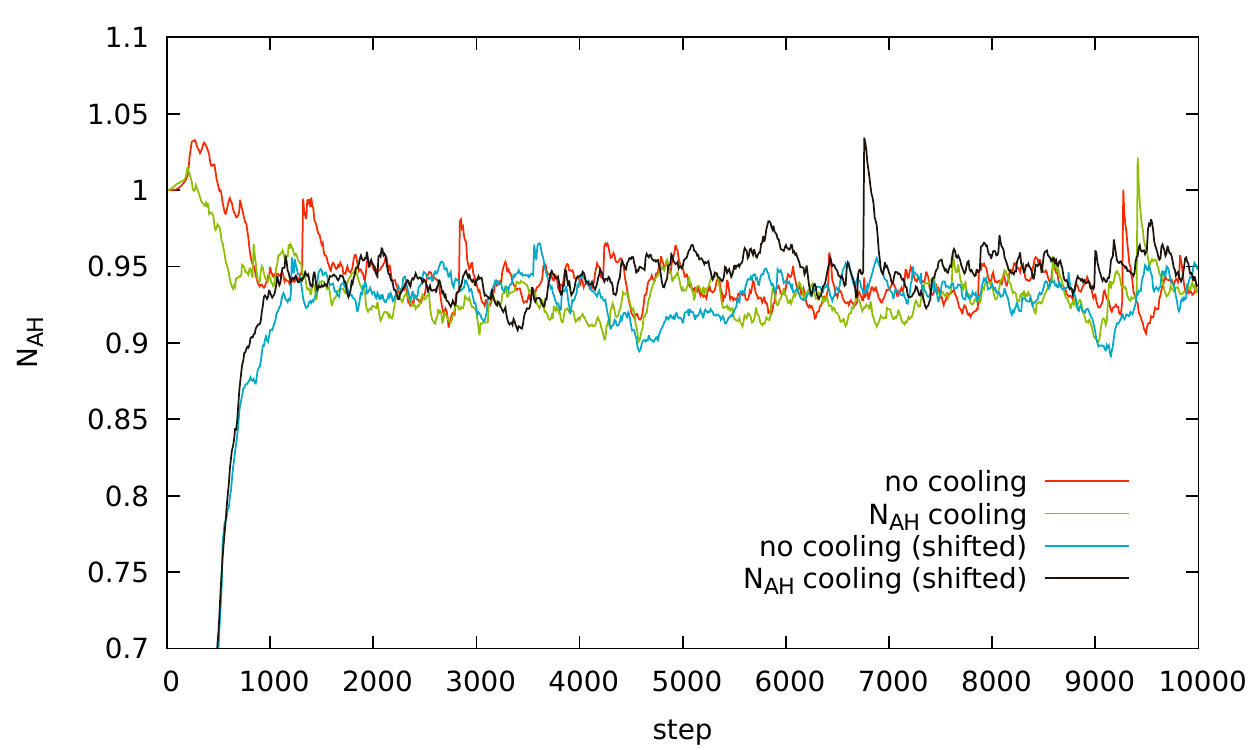}
    \includegraphics[width=0.45\textwidth]{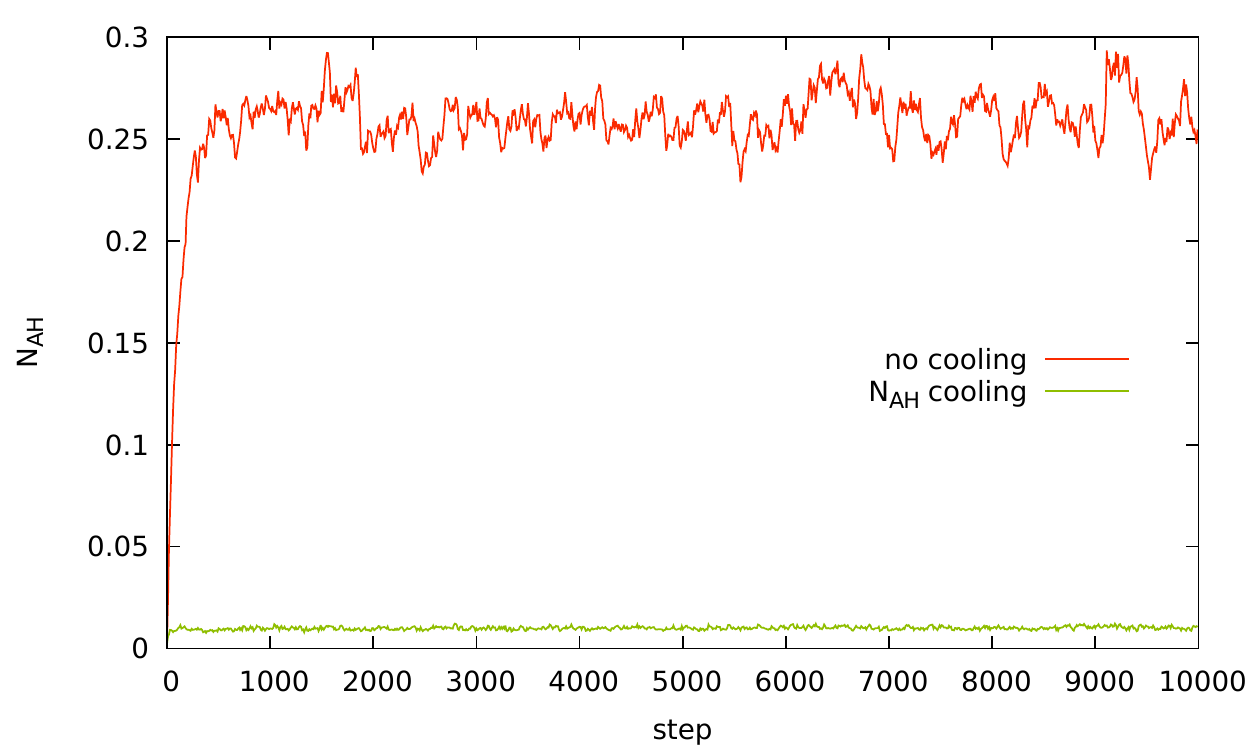}
  \end{center}
  \caption{Value of the anti-hermiticity norm, $\mathcal{N}_{AH}$, with (green curves) and without cooling
    (red curves) as a function of Langevin time.
    Stephanov model (LHS), Osborn model (RHS). The Stephanov plot
    also includes the values from the shifted representation with (purple curve)
    and without cooling (blue curve). These start at 0
    for $t = 0$, but very quickly join the unshifted curves.}
  \label{fig:ah_cooling_norm}
\end{figure}

\begin{figure}
  \begin{center}
    \includegraphics[width=0.45\textwidth]{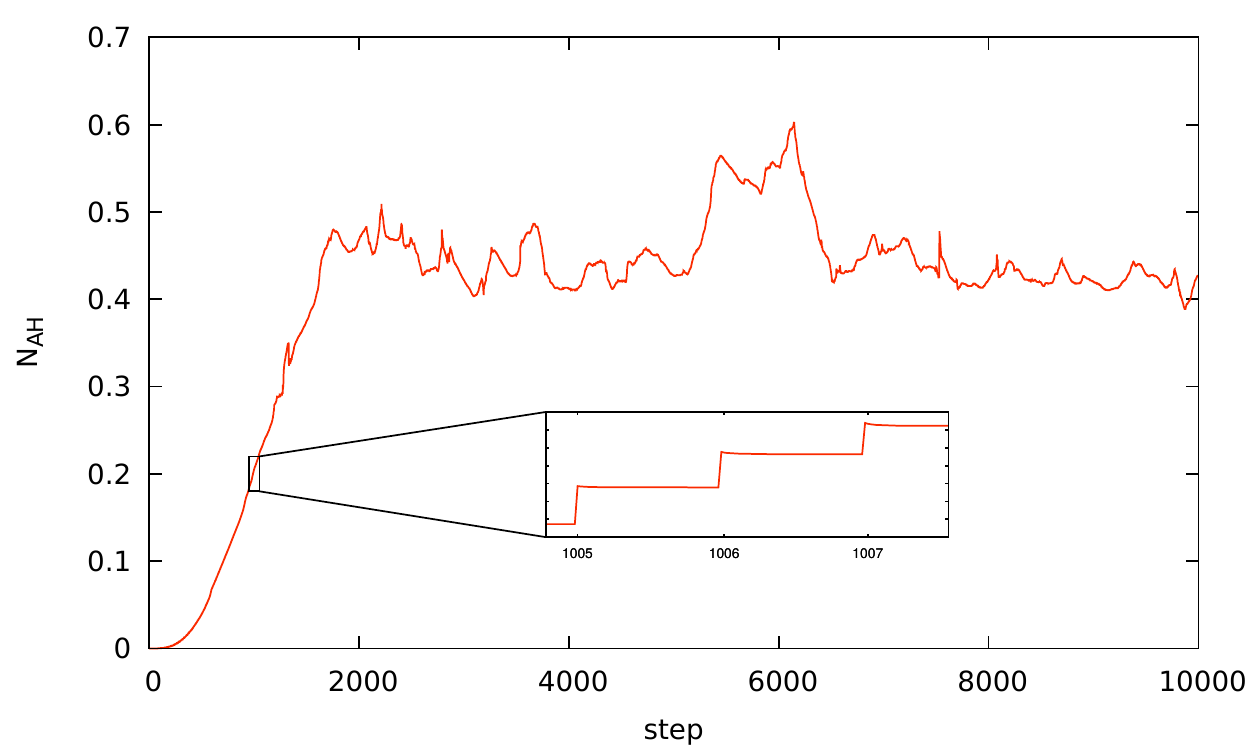}
    \includegraphics[width=0.45\textwidth]{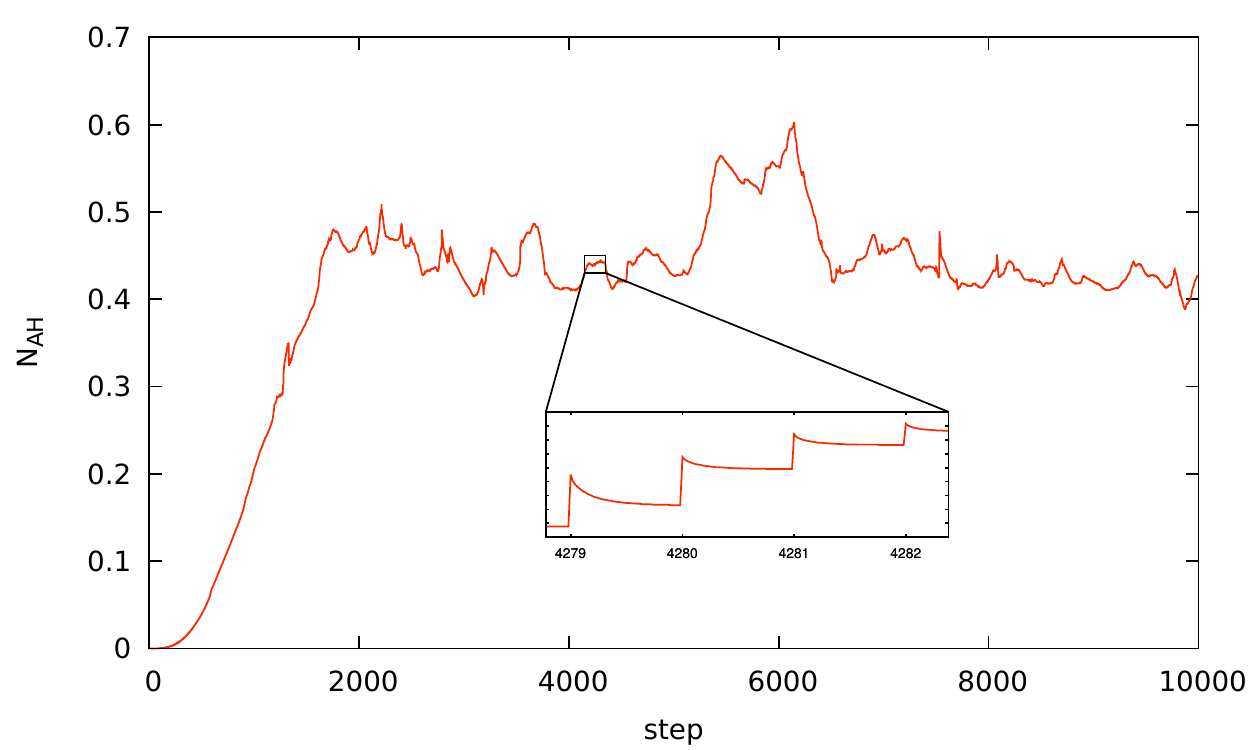}
  \end{center}
  \caption{The anti-Hermiticity norm, $\mathcal{N}_{AH}$, for the shifted representation of the Stephanov
    model as a function of Langevin time for $8\times{}8$ block matrices. The
    magnification in the inset 
     shows the effect of cooling  on $\mathcal{N}_{AH}$.
    In this specific analysis, we use 50 gauge cooling transformations
    between each Langevin step.}
  \label{fig:ah_cooling_spy_plots}
\end{figure}

We now focus on $\mathcal{N}_{AH}$ (figure \ref{fig:eval_cooling_norm}) and
$\mathcal{N}_{\mathrm{ev}}$ (figure \ref{fig:ah_cooling_norm}) as a function of
Langevin time. We observe that in the case of the Osborn model the norms are
heavily reduced by the application of cooling but the Stephanov model is
unaffected. Even in the extreme case of the shifted representation (see the next
section), which through its advantageous initial conditions could have been a
promising fix, we see that we obtain results similar to those without cooling and/or
shifting. In figure \ref{fig:ah_cooling_spy_plots} we show our results for
$N=8$ matrices where we plot the norm and zoom in, in order to see better the
effects of cooling and Langevin updates.  Once thermalization has been reached
we see that the cooling does not have a great enough impact on
the simulation as to overcome a single Langevin update; this does not change
when decreasing the CL algorithm's stepsize, indicating that the solutions we
are interested in are out of reach of the $Gl(N)$ transformations.

\section{Shifted representation}

Another possible resolution of the problematic issues of the algorithm is to shift the chemical potential out of the fermion determinant to the "gauge" degrees of freedom. This can be achieved by a simple change of variables. The Dirac operator initially reads
\begin{equation}
  D = %
  \begin{pmatrix}
    m & iA - B + \mu \\
    iA^T + B^T + \mu & m
  \end{pmatrix}.
\end{equation}
One can absorb $\mu$ into $A$ as follows, $A' = A - i\mu$. In terms of the
matrices $A'$ and $B$ the action reads
\begin{equation}
  S = N \tr \big( A'^T A' + 2i\mu A' - \mu^2 + B^T B\big) %
    - N_f \tr \log \big(m^2 + X' Y'\big),
\end{equation}
where $X' = A' + iB$ and $Y' = A'^T - iB^T$. This result in different
Langevin forces~\cite{CLarticle}.

\begin{figure}
  \begin{center}
    \includegraphics[width=0.45\textwidth]{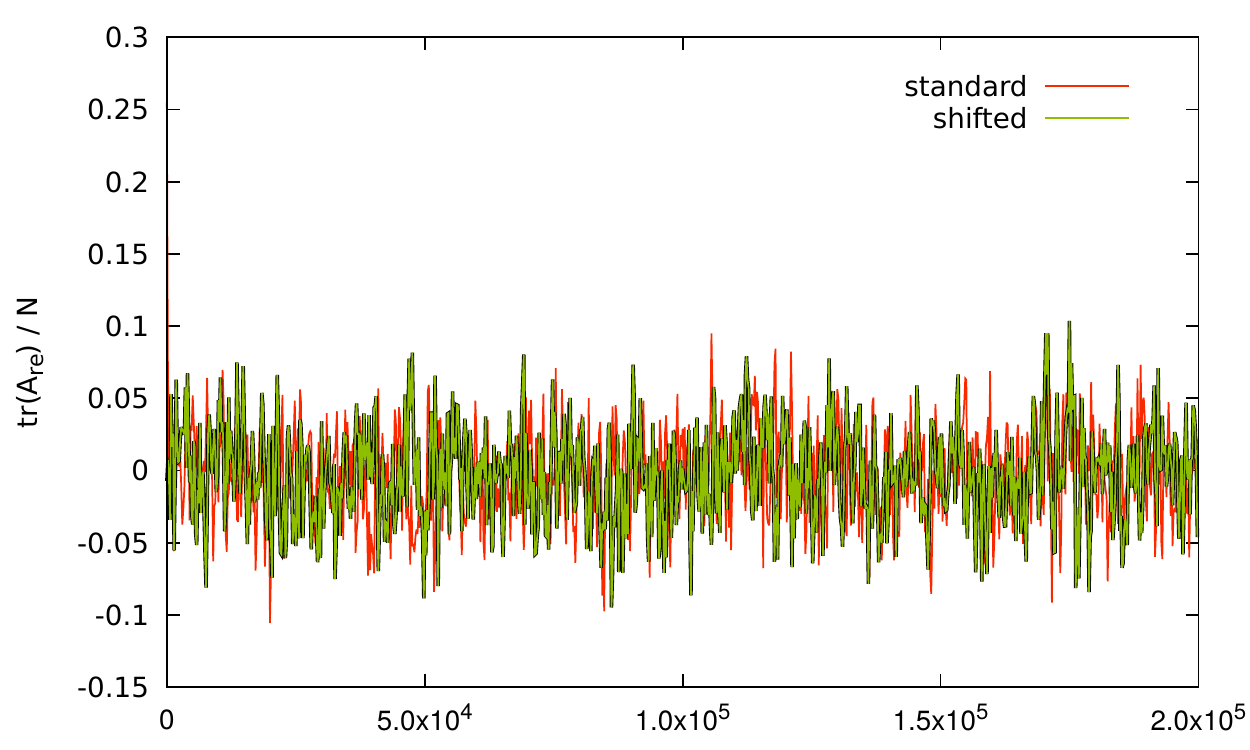}
    \includegraphics[width=0.45\textwidth]{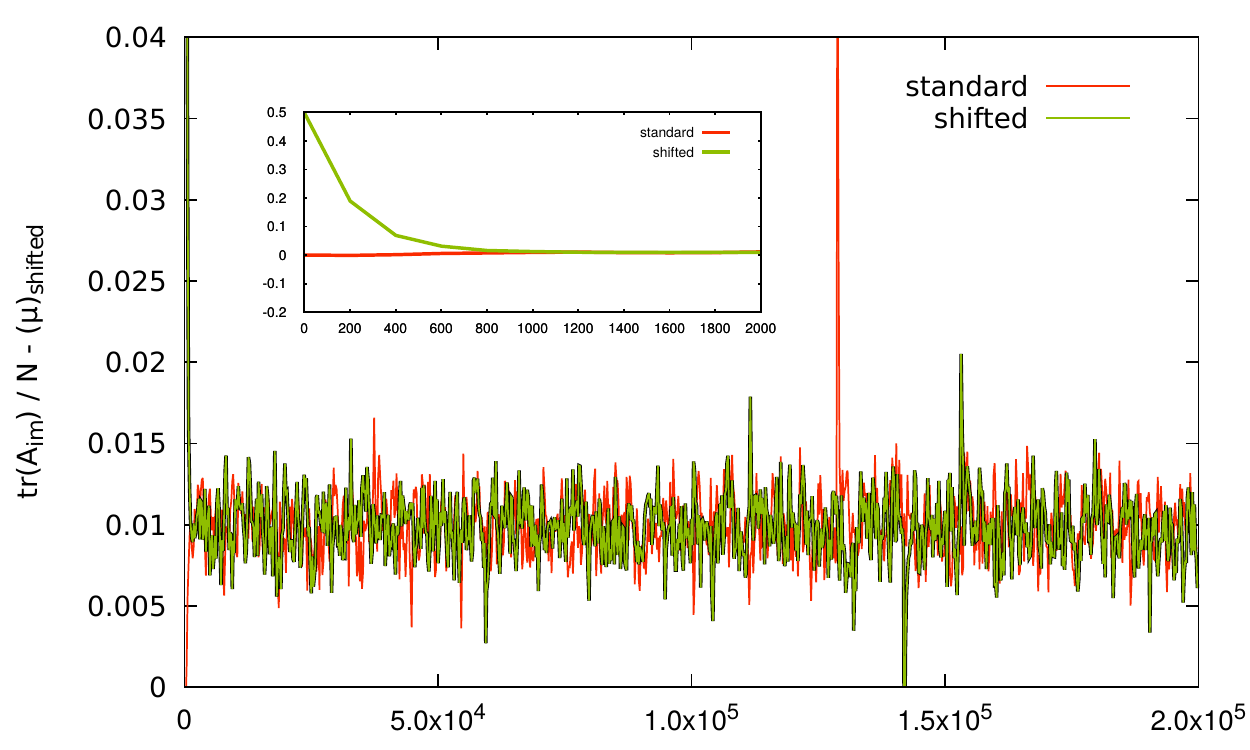}
  \end{center}
  \caption{Average diagonal entry of $A$ (red curves) and $A'+i\mu$ (green curves)
    as a function of the algorithmic
    steps. The real
    component of the trace is shown on the left and the imaginary component on the
    right. The inset in the right figure shows the first 2000 Langevin steps.}
  
  \label{fig:shifted_diag_average}
\end{figure}

To analyze the dynamics of the shifted representation we analyze the matrix
elements of
 $A$ and $A'$ during a typical CL simulation; these are shown in
figure \ref{fig:shifted_diag_average}. Although the two matrices start out very
different, they do coincide after thermalization. This means that
\begin{equation}
  \big\langle A' \big\rangle_{\mathrm{CL},\text{shifted}} = \big\langle A
  \big\rangle_{\mathrm{CL},\text{standard}} - i \mu,
\end{equation}
and thus they converge to the same solution. 

\section{Conclusions}
In this contribution we have performed a detailed comparison of the effects of gauge cooling on two different matrix models for QCD at finite density. We have shown that gauge cooling can fix the pathologies of the Osborn model, where the sign problem is artificial and can even be removed by careful redefinition of the parameters, but fails to overcome the problems of the Stephanov model which is a true matrix model for QCD at finite baryon density. The Stephanov model, possessing a phase transition to a phase with non-zero baryon density, is an excellent candidate for this type of investigations since dedicated studies have shown that CL fails when one approaches the deconfinement transition region of the QCD phase diagram, cf.\ \cite{Fodor} and \cite{Bloch17} for an attempt to fix this problem. 
\vspace*{-0.15cm}

\section{Acknowledgements}
{\small SZ acknowledges
support by the National Science Foundation (USA) under
grant PHY-1516509 and by the Jefferson Science Associates,
LLC under U.S. DOE Contract \# DE-AC05-
06OR23177. JV acknowledges partial support from U.S. DOE Grant No. DEFAG-
88FR40388. JB was supported by the Deutsche Forschungsgemeinschaft
(SFB/TRR-55). The authors would like to thank G. Aarts, I. O. Stamatescu, K. Nagata and J. Nishimura for fruitful discussions.}
\vspace*{-0.15cm}

\bibliography{lattice2017}

\end{document}